\documentclass[a4paper,11pt]{article}
\usepackage{jheppub}
\usepackage{mathtools}

\graphicspath{{plots/}}

\newcommand{\gs}{g_\star}
\newcommand{\gss}{g_{\star s}}
\newcommand{\Hrh}{H_\text{rh}}
\newcommand{\Trh}{T_\text{rh}}
\newcommand{\Tmax}{T_\text{max}}
\newcommand{\Tfin}{T_\text{fin}}
\newcommand{\Tini}{T_\text{ini}}
\newcommand{\arh}{a_\text{rh}}
\newcommand{\afin}{a_\text{fin}}
\newcommand{\aini}{a_\text{ini}}
\newcommand{\kfin}{k_\text{fin}}
\newcommand{\krh}{k_\text{rh}}
\newcommand{\rgw}{\rho_\text{GW}}
\newcommand{\rp}{\rho_\phi}
\newcommand{\rR}{\rho_R}
\newcommand{\DNeff}{\Delta N_\text{eff}}
\newcommand{\TDNeff}{T_{\Delta N_\text{eff}}}
\newcommand{\fpeak}{f_\text{peak}}

\title{Thermal Gravitational Waves During Reheating}
\author[a]{Nicolás Bernal}
\author[b]{and Yong Xu}
\affiliation[a]{New York University Abu Dhabi\\
	PO Box 129188, Saadiyat Island, Abu Dhabi, United Arab Emirates}
\affiliation[b]{PRISMA$^+$ Cluster of Excellence and Mainz Institute for Theoretical Physics\\
	Johannes Gutenberg University, 55099 Mainz, Germany}
\emailAdd{nicolas.bernal@nyu.edu}
\emailAdd{yonxu@uni-mainz.de}

\abstract{In this work, we revisit the generation of stochastic gravitational waves (GWs) from interactions in the thermal plasma. We extend the existing literature by incorporating the reheating phase into the thermal history. Our results show that the amplitude of the GW spectrum can be significantly enhanced because the temperature during reheating can be much higher than the reheating temperature. Furthermore, since the temperature evolution during reheating differs from that of free radiation, the peak frequency of the spectrum can also shift. Additionally, the morphology of the spectrum can present characteristic features. We also compute the contribution of the integrated GW spectrum to the effective number of neutrino species, $\DNeff$, which can be substantially larger.}
\begin{document}
	\begin{flushright}
		MITP-24-076
	\end{flushright}
	\maketitle
	
	%%%%%%%%%%%%%%%%%%%%%%%%%%%%%%%%%%%%%
	\section{Introduction}
	%%%%%%%%%%%%%%%%%%%%%%%%%%%%%%%%%%%%%
	The precise measurement of light element abundances implies that the Universe must have reached thermal equilibrium at a temperature of at least $T \sim \mathcal{O}(\text{MeV})$ to have a successful Big Bang nucleosynthesis (BBN)~\cite{Sarkar:1995dd, Kawasaki:2000en, Hannestad:2004px, deSalas:2015glj, Cyburt:2015mya, Hasegawa:2019jsa}. Due to unavoidable gravitational couplings, a stochastic background of gravitational waves (GWs) is generated from graviton production through interactions of standard model (SM) particles in the thermal plasma~\cite{Ghiglieri:2015nfa, Ghiglieri:2020mhm, Ringwald:2020ist, Klose:2022knn, Klose:2022rxh, Ringwald:2022xif, Ghiglieri:2022rfp, Drewes:2023oxg, Ghiglieri:2024ghm}. Since the energy of these gravitons redshifts in the same way as the temperature in a radiation-dominated phase, the resulting GW spectrum would resemble that of the cosmic microwave background (CMB), peaking at a frequency of $f_\text{peak} \sim 80$~GHz. Furthermore, the rate of graviton production scales as $\propto T^3/M_P^2$~\cite{Ghiglieri:2015nfa, Ghiglieri:2020mhm, Ringwald:2020ist}, with $M_P$ being the Planck mass, implying that the amplitude of the GW spectrum is controlled by the highest temperatures reached by the radiation bath. In the radiation-dominated era, this highest temperature corresponds to the reheating temperature $\Trh$.
	
	In this work, we revisit the generation of GWs from the thermal plasma, aiming to extend previous analyses~\cite{Ghiglieri:2015nfa, Ghiglieri:2020mhm, Ringwald:2020ist, Klose:2022knn, Klose:2022rxh, Ringwald:2022xif, Ghiglieri:2022rfp, Drewes:2023oxg, Ghiglieri:2024ghm} by incorporating the thermal history beyond the reheating temperature $\Trh$\footnote{We note that the temperatures under consideration are well below the Planck scale, so the graviton would not thermalize. For an alternative scenario, see Ref.~\cite{Vagnozzi:2022qmc}.}. This extension is necessary for the following reason. It is well-known that the standard thermal Big Bang theory is incomplete. To resolve issues such as flatness, horizon, and monopole problems, an inflationary phase is strongly motivated~\cite{Starobinsky:1980te, Guth:1980zm, Linde:1981mu, Albrecht:1982wi}. However, the exponentially-fast expansion during cosmic inflation ends with an extremely cold Universe. In order to transition to a background dominated by thermal radiation, the Universe must undergo a reheating phase~\cite{Kofman:1997yn}. During reheating, the maximum SM temperature $\Tmax$ reached by the plasma could be much higher than the reheating temperature $\Trh$~\cite{Giudice:2000ex}. This suggests that graviton production during the reheating phase could be significantly more efficient than in the subsequent radiation-dominated phase.
	
	The purpose of this study is to account for the reheating phase when calculating the GW background from thermal plasma, which is currently lacking in the literature.\footnote{We note that Ref.~\cite{Muia:2023wru, Villa:2024jkw} studies thermal GW during a kination phase.} As mentioned above, the temperature during reheating can exceed the reheating temperature. We anticipate that this will lead to an enhancement in the peak of the GW spectrum. Additionally, depending on the temperature scaling during reheating, the location of the peak frequency can also change, generating new spectral features. Consequently, the GW spectrum produced during reheating is expected to differ from that in the radiation phase.\footnote{Other sources of GWs during reheating count non-perturbative preheating effects~\cite{Caprini:2018mtu}, and graviton bremsstrahlung or pair production from inflaton decays or scatterings~\cite{Ema:2015dka, Ema:2016hlw, Nakayama:2018ptw, Huang:2019lgd, Ema:2020ggo, Ema:2021fdz, Barman:2023ymn, Barman:2023rpg, Kanemura:2023pnv, Bernal:2023wus, Tokareva:2023mrt, Choi:2024ilx, Hu:2024awd, Barman:2024htg, Xu:2024fjl, Inui:2024wgj,  Jiang:2024akb}; however, contrary to the present scenario, they depend on the particular properties of the inflaton. GWs from bremsstrahlung could also be sourced from particles other than inflatons~\cite{Choi:2024acs, Hu:2024bha, Datta:2024tne}. }
	
	In this work, we parameterize the reheating phase using two key quantities: the equation-of-state parameter $\omega$ of the Universe during reheating and the temperature scaling parameter $\alpha$, where $T \sim a^{-\alpha}$ and $a$ denotes the cosmic scale factor. We analyze the resulting GW spectrum for general values of $\omega$ and $\alpha$, comparing it to the standard GW spectrum produced during the radiation-dominated era. Given that GWs contribute to dark radiation, we further calculate their impact on the effective number of neutrino species $\DNeff$.
	
	The paper is organized as follows. In Section~\ref{sec:framework}, we offer the setup for computing thermal GWs. In Section~\ref{eq:reheating}, we revisit reheating with a focus on parameterization of the background. We investigate $\DNeff$ in Section~\ref{sec:DNeff}, in particular the contribution from the reheating phase.  In Section~\ref{sec:TGW}, we present the GW spectrum with a focus on the contribution of the reheating phase. Finally, we summarize our main findings in Section~\ref{sec:concl}.
	
	%%%%%%%%%%%%%%%%%%%%%%%%%%%%%%%%%%%%%
	\section{The Set-up} \label{sec:framework}
	%%%%%%%%%%%%%%%%%%%%%%%%%%%%%%%%%%%%%
	We begin with the evolution of the GW energy density $\rgw$, which is governed by the Boltzmann equation
	\begin{equation} \label{eq:BE0}
		\frac{d\rgw}{dt} + 4\, H\, \rgw = \gamma\,,
	\end{equation}
	where $H$ denotes the Hubble parameter, $t$ the cosmic time, and $\gamma$ corresponds to the graviton production rate density from the SM thermal bath. It takes a form~\cite{Ghiglieri:2020mhm,  Ringwald:2020ist}
	\begin{equation} \label{eq:gamma1}
		\gamma(T) = \frac{4\, T^4}{M_P^2} \int \frac{d^3\vec k}{(2\pi)^3}\, \hat\eta(T, \hat k) = \frac{2\, T^4}{\pi^2\, M_P^2} \int dk\, k^2\, \hat\eta(T, \hat k),
	\end{equation}
	with $M_P \simeq 2.4 \times 10^{18}$~GeV being the reduced Planck mass, and the dimensionless parameter $\hat k \equiv k/T$ defined as a function of the graviton momentum $k$. For later convenience to compute the GW spectrum, we also write the differential form of the rate density as
	\begin{equation} \label{eq:diff}
		\frac{d\gamma}{dk}(T, \hat k) = \frac{2\, T^6}{\pi^2\, M_P^2}\, \hat k^2\, \hat\eta(T, \hat k).
	\end{equation}
	The source term $\gamma$ encodes all contributions from SM particles to graviton production. For temperatures above the electroweak crossover, it is given by~\cite{Ghiglieri:2020mhm, Ringwald:2020ist}
	\begin{equation} \label{eq:etahat}
		\hat\eta(T, \hat k) \simeq
		\begin{dcases}
			\frac{\bar\eta}{g_1^4(T)\, \ln\left[\frac{5}{\hat m_1(T)}\right]} & \text{ for } \hat k \lesssim \alpha_1^2\,,\\
			\hat\eta_\text{HTL}(T, \hat k) + \hat\eta_2(T, \hat k) & \text{ for } \hat k \gtrsim \max(\hat m_n)\,,
		\end{dcases}
	\end{equation}
	where $\bar\eta$ is~\cite{Arnold:2000dr} 
	\begin{equation}
		\bar\eta = \zeta(5)^2 \left(\frac52\right)^3 \left(\frac{12}{\pi}\right)^5 \frac{3}{2(1232+9\pi^2)} \simeq 15.51\,.
	\end{equation}
	We note that the first piece in Eq.~\eqref{eq:etahat} comes from the hydrodynamic contribution for momenta lower than the temperature. In the large-momenta regime, the contributions are from microscopic particle scatterings. The dominant contributions arise from hard thermal loops (HTL), and is given by~\cite{Ghiglieri:2020mhm, Ringwald:2020ist}
	\begin{equation}\label{eq:HTL}
		\hat\eta_\text{HTL}(T, \hat k) = \frac{\hat k}{16\pi\, (e^{\hat k}-1)} \sum_{n=1}^3 N_n\, \hat m_n^2(T)\, \log\left[1 + \frac{4\, \hat k^2}{\hat m_n^2(T)}\right].
	\end{equation}
	The other terms in the second part of Eq.~\eqref{eq:etahat} is~\cite{Ringwald:2020ist}
	\begin{align}
		\hat\eta_2(T, \hat k) &\equiv \left(3 g_2^2 +12 g_3^2\right) \eta_{gg}(\hat k) + \left(g_1^2 + 3 g_2^2\right) \eta_{sg}(\hat k) \nonumber\\
		&\quad + \left(5 g_1^2 + 9 g_2^2 + 24 g_3^2\right) \eta_{fg}(\hat k) + \left(3 |y_t|^2 + 3 |y_b|^2 + |y_\tau|^2\right) \eta_{sf}(\hat k)\,,
	\end{align}
	arising from infrared-finite loop integrals involving only gauge fields $\eta_{gg}$, scalars and gauge fields $\eta_{sg}$, fermions and gauge fields $\eta_{fg}$, and scalars and fermions $\eta_{sf}$. Here, $y_t$, $y_b$, and $y_\tau$ denote the Yukawa couplings for the top quark, the bottom quark, and the tau lepton, respectively. The thermal mass parameters $\hat m_i(T)$ are defined via
	\begin{equation}
		\hat m_i^2(T) \equiv \frac{m_i^2(T)}{T^2} = C_i\, g_i^2(T),
	\end{equation}
	with $C_1 = C_2 = 11/6$, $C_3 = 2$. Finally, $N_1=1$, $N_2=3$, and $N_3=8$ are the number of generators of $U(1)_Y$, $SU(2)_L$ and $SU(3)_c$, respectively, and $g_1(T)$, $g_2(T)$, $g_3(T)$ are the values of the gauge coupling strengths. We note that after integrating Eq.~\eqref{eq:gamma1}, one arrives at
	\begin{equation}\label{eq:scaling}
		\gamma(T) \simeq \mathcal{C}\, \frac{T^7}{M_P^2}\,,
	\end{equation}
	with $\mathcal{C} \simeq 0.73$, for the integrated GW production rate. 
	
	%%%%%%%%%%%%%%%%%%%%%%%%%%%%%%%%%%%%%%%%%%%%%%%%%%%%%%%%%%
	\section{Parametrizing Cosmic Reheating} \label{eq:reheating}
	%%%%%%%%%%%%%%%%%%%%%%%%%%%%%%%%%%%%%%%%%%%%%%%%%%%%%%%%%%
	The detailed dynamics of the reheating phase remains uncertain, but is typically assumed to be driven by the decay or annihilation of a scalar field $\phi$, such as the inflaton, characterized by an effective equation-of-state parameter $\omega$. This implies that the energy density of the scalar field scales as $\rho_\phi(a) \propto a^{-3(1+\omega)}$ during reheating. To reheat the Universe, energy must be transferred from $\phi$ to the SM radiation, with a corresponding energy density given by
	\begin{equation}
		\rR(T) \equiv \frac{\pi^2}{30}\, \gs(T)\, T^4,
	\end{equation}
	where where $\gs(T)$ is the effective number of relativistic degrees of freedom contributing to the radiation energy. The reheating temperature $\Trh$ is defined as the temperature at which the radiation energy density is equal to the inflaton energy density, $\rp (\arh) = \rR (\arh) = 3\, \Hrh^2\, M_P^2$, where $\arh$ is the scale factor at $T = \Trh$, and $\Hrh \equiv H(\arh)$ is the Hubble expansion rate at that time. It follows that the Hubble rate is~\cite{Bernal:2024yhu, Bernal:2024ndy}
	\begin{equation} \label{eq:Hubble}
		H(a) = \Hrh \times
		\begin{dcases}
			\left(\frac{\arh}{a}\right)^\frac{3(1+\omega)}{2} &\text{ for } a \leq \arh,\\
			\left(\frac{\gs(T)}{\gs(\Trh)}\right)^\frac12 \left(\frac{\gss(\Trh)}{\gss(T)}\right)^\frac23 \left(\frac{\arh}{a}\right)^2 &\text{ for } \arh \leq a\,,
		\end{dcases}
	\end{equation}
	where $\gss(T)$ denotes the effective number of relativistic degrees of freedom contributing to the entropy density
	\begin{equation}
		s(T) = \frac{2 \pi^2}{45}\, \gss(T)\, T^3.
	\end{equation}
	Additionally, the evolution of the SM temperature during reheating is parametrized as $T(a) \propto a^{-\alpha}$, and therefore~\cite{Bernal:2024yhu, Bernal:2024ndy}
	\begin{equation}\label{eq:Tem}
		T(a) = \Trh \times
		\begin{dcases}
			\left(\frac{\arh}{a}\right)^\alpha &\text{ for } \aini \leq a \leq \arh,\\
			\left(\frac{\gss(\Trh)}{\gss(T)}\right)^\frac13 \frac{\arh}{a} &\text{ for } \arh \leq a\,,
		\end{dcases}
	\end{equation}
	where $\Tini \equiv T(\aini) =  \Trh \times (\arh/\aini)^\alpha$. After reheating, the Universe transitions to a radiation-dominated phase, where $H(a) \propto a^{-2}$ and $T(a) \propto a^{-1}$, as expected.
	
	%%%%%%%%%%%%%%%%%%%%%%%%%%%%%%%%%%%%%%%%%%%%%%%%%%%
	\begin{figure}[t!]
		\def\sepf{0.496}
		\centering
		\includegraphics[width=\sepf\columnwidth]{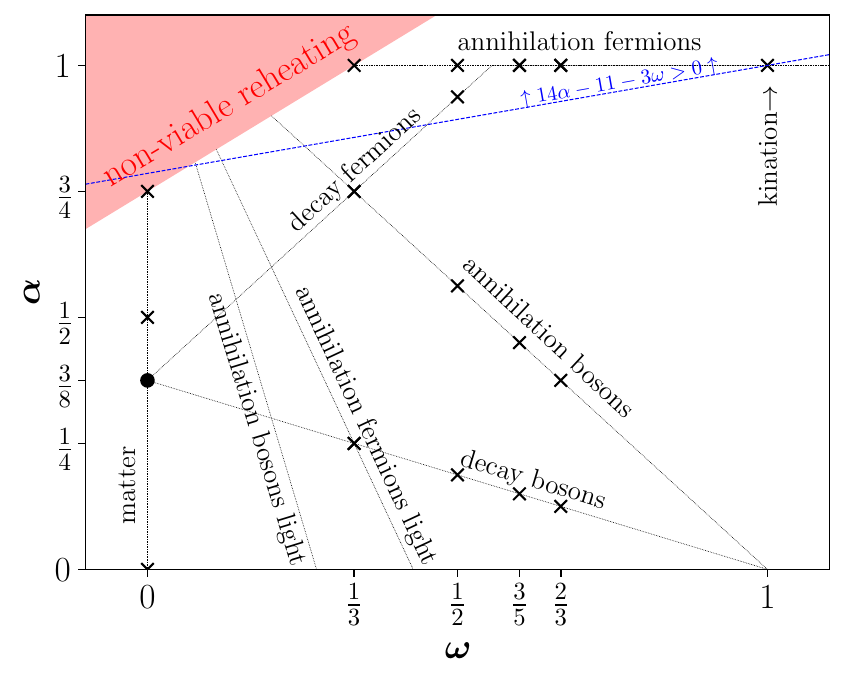}
		\caption{Summary of the different reheating scenarios. The black dot corresponds to the standard case where the inflaton scales as non-relativistic matter and decays into SM particles with a constant decay width, while the black crosses correspond to the alternative scenarios described in the text. The red area in the upper left corner does not give rise to viable reheating. Above, over and below the dashed blue line, the GW spectrum has a power-law, logarithmic or order-one enhancement during reheating, respectively; cf. Eq~\eqref{eq:DNeff1}.}
		\label{fig:alpha_omega}
	\end{figure} 
	%%%%%%%%%%%%%%%%%%%%%%%%%%%%%%%%%%%%%%%%%% 
	Here, we offer a few comments about the way reheating is parametrized. The most popular reheating scenario corresponds to a heavy inflaton with an energy density that scales as non-relativistic matter ($\omega = 0$), which perturbatively decays into a couple of SM particles, giving rise to a scaling of the temperature of $\alpha = 3/8$~\cite{Giudice:2000ex}. This scenario can be realized in viable inflationary models such as Starobinsky inflation~\cite{Starobinsky:1980te} or polynomial inflation~\cite{Drees:2021wgd, Bernal:2021qrl, Drees:2022aea, Bernal:2024ykj}.  However, other scenarios are also possible. For example, within the framework of the $\alpha$-attractor inflation model~\cite{Kallosh:2013hoa, Kallosh:2013maa}, it is possible that the inflaton oscillates at the bottom of a monomial potential $V(\phi) \propto \phi^n$ with $n\geq 2$ during reheating, with a corresponding equation-of-state parameter $\omega = (n-2)/(n+2)$~\cite{Turner:1983he}. Furthermore, the value of $\alpha$ depends on the mechanism by which the inflaton energy is transferred to the radiation, specifically the nature of the inflaton-matter couplings~\cite{Co:2020xaf, Garcia:2020wiy, Xu:2023lxw, Barman:2024mqo}. For example, when the inflaton decays into bosons, one has $\alpha = 3/(2(n+2))$, and when it decays into fermions with $n < 7$, $\alpha = 3(n-1)/(2(n+2))$. In the case of fermionic inflaton decay and $n > 7$, one has $\alpha = 1$. When considering inflaton annihilation into bosons through contact interactions, $\alpha = 9/(2(n+2))$ for $n \geq 3$. In the case of reheating via $s$-channel annihilation mediated by a light scalar, resonant effects can lead to $\alpha = 3\, (7-2n)/(2(n+2))$ and $\alpha = 3\, (5-n)/(2(n+2))$ for bosonic and fermionic final states, respectively~\cite{Barman:2024mqo}. If the mediator is massive, one has $\alpha=1$ for inflaton annihilations into a pair of fermions~\cite{Barman:2024mqo}. Moreover, it is also possible to have a constant temperature during reheating with $\alpha =0$~\cite{Co:2020xaf, Barman:2022tzk, Chowdhury:2023jft, Cosme:2024ndc}. If the inflaton energy density is diluted faster than free radiation, that is, if $\omega > 1/3$, it is not necessary for the inflaton to decay or annihilate away, and then one can have $\alpha = 1$, as in the case of kination~\cite{Spokoiny:1993kt, Ferreira:1997hj}. Finally, it is possible to have $\alpha>1$, in particular $\alpha=2$ for non-relativistic particles in kinetic equilibrium.
	
	The different reheating scenarios mentioned above are shown in Fig.~\ref{fig:alpha_omega}, in the plane [$\omega$,  $\alpha$]. The black lines represent specific reheating scenarios, which illustrates the relationship between these parameters and the dynamics of the reheating process. The vertical gray dotted line corresponds to a scenario with $\omega=0$. The blue dotted line represents $14\alpha - 11 -3\omega =0$, a special case that will be frequently used in later discussions.  The red area with $\alpha \leq 3(1+\omega)/4$ in the upper left corner does not give rise to viable reheating, as for these parameters, the SM radiation energy density never overcomes the inflaton energy density. 
	
	%%%%%%%%%%%%%%%%%%%%%%%%%%%%%%%%%%%%%
	\section{\boldmath $\DNeff$} \label{sec:DNeff}
	%%%%%%%%%%%%%%%%%%%%%%%%%%%%%%%%%%%%%
	Since the GW energy density scales as radiation, in this section we explore their contribution to the effective number of neutrino species $\DNeff$. The effective number of neutrinos, $N_\text{eff}$, is defined based on the total radiation energy density in the late Universe at a photon temperature $\TDNeff$, expressed as
	\begin{equation}
		\rho_\text{rad}(\TDNeff) = \rho_\gamma + \rho_\nu + \rgw = \left[1 + \frac78 \left(\frac{T_\nu}{T_\gamma}\right)^4 N_\text{eff}\right] \rho_\gamma(\TDNeff)\,.
	\end{equation}
	Here, $\rho_\gamma$, $\rho_\nu$, and $\rho_{\text{GW}}$ denote the energy densities of photons, SM neutrinos, and GWs, respectively. The ratio of neutrino-photon temperatures after neutrino decoupling is given by $T_\nu / T_\gamma = (4/11)^{1/3}$. Within the SM, the prediction for the effective number of neutrinos, which accounts for the non-instantaneous decoupling of neutrinos, is $N_\text{eff}^\text{SM} = 3.044$~\cite{Dodelson:1992km, Hannestad:1995rs, Dolgov:1997mb, Mangano:2005cc, deSalas:2016ztq, EscuderoAbenza:2020cmq, Akita:2020szl, Froustey:2020mcq, Bennett:2020zkv}. Taking the contribution from GWs, it follows that~\cite{Barman:2023ymn}
	\begin{equation} \label{eq:DNeff}
		\DNeff \equiv N_\text{eff} - N_\text{eff}^\text{SM} = \frac87 \left(\frac{11}{4}\right)^\frac43 \frac{\rgw(T_{\DNeff})}{\rho_\gamma(T_{\DNeff})} = \frac87 \left(\frac{11}{4} \frac{\gss(T_{\DNeff})}{\gss(\Tfin)}\right)^\frac43 \frac{\gs(\Tfin)}{2} \frac{\rgw(\Tfin)}{\rR(\Tfin)},
	\end{equation}
	where $\Tfin$ denotes the temperature at which the GWs from thermal plasma are evaluated. Following Ref.~\cite{Ringwald:2022xif}, we take $\Tfin$ to be the temperature at which the electroweak crossover takes place.
	
	%%%%%%%%%%%%%%%%%%%%%%%%%%%%%%%%%%%%
	\begin{table}[t!]
		\begin{center}
			\begin{tabular}{|c||c|}
				\hline
				$\boldsymbol{\Delta N_\text{\bf eff}}$ & {\bf Experiments}  \\ 
				\hline\hline
				$0.34$ & Planck 2018~\cite{Planck:2018vyg}\\
				\hline
				$0.14$ & BBN+CMB combined~\cite{Yeh:2022heq}\\
				\hline\hline
				$0.06$ & CMB-S4~\cite{Abazajian:2019eic}, PICO~\cite{NASAPICO:2019thw} \\
				\hline
				$0.027$ & CMB-HD~\cite{CMB-HD:2022bsz}\\
				\hline
				$0.013$ &  COrE~\cite{COrE:2011bfs}, Euclid~\cite{EUCLID:2011zbd} \\
				\hline\hline
				$3\times 10^{-6}$ &  CVL~\cite{Ben-Dayan:2019gll}\\
				\hline
			\end{tabular}
		\end{center}
		\caption {Present constraints and projections on $\DNeff$ from different experiments. CVL corresponds to hypothetical limit from cosmic variance.}
		\label{tab:DNeff}
	\end{table}
	%%%%%%%%%%%%%%%%%%%%%%%%%%%%%%%%%%%%%
	Experimental bounds on $\DNeff$ are provided for two key epochs: $\TDNeff = T_\text{BBN}$, corresponding to the time of BBN, and $\TDNeff = T_\text{CMB}$, when photons decouple from the thermal plasma during recombination. There are several bounds (Planck~\cite{Planck:2018vyg} and BBN+CMB combined data~\cite{Yeh:2022heq}) and projections (CMB-S4~\cite{Abazajian:2019eic}, PICO~\cite{NASAPICO:2019thw}, CMB-HD~\cite{CMB-HD:2022bsz}, COrE~\cite{COrE:2011bfs} and Euclid~\cite{EUCLID:2011zbd}) on $\DNeff$. Furthermore, as mentioned in Ref.~\cite{Ben-Dayan:2019gll}, a hypothetical cosmic-variance-limited (CVL) CMB polarization experiment could presumably be reduced to as low as $\DNeff \simeq 3 \times 10^{-6}$. We briefly summarize them in Table~\ref{tab:DNeff}. 
	
	Now, our task is to compute $\rgw(\Tfin)$ by solving Eq.~\eqref{eq:BE0}. Since we will also consider the contribution of graviton production during reheating to the total energy density of GW, it is convenient to rewrite Eq.~\eqref{eq:BE0} with the scale factor as variable. This leads to
	\begin{equation} \label{eq:BE1}
		\frac{d(a^4\, \rgw)}{da}  = \frac{a^3}{H(a)}\, \gamma(T)\,,
	\end{equation}
	and therefore $\rgw(\Tfin)$ is given by
	\begin{equation} \label{eq:BE2}
		\rgw(\afin) = \frac{1}{\afin^4} \int_{\aini}^{\afin} \frac{a^3}{H(a)}\, \gamma(T)\, da\,,
	\end{equation}
	where $\aini$ denotes the scale factor at the onset of the radiation bath {\it during reheating}, and $\afin$ corresponds to the scale factor when $T = \Tfin$. As mentioned above, by incorporating the reheating phase, it follows that $\aini < \arh < \afin$. To our knowledge, the regime $\aini < a < \arh$ has not been previously considered in the literature.\footnote{In the context of string cosmology, the GW spectrum generated by the gas of long strings was studied in Refs.~\cite{Frey:2024jqy, Villa:2024jbf}.}
	
	%%%%%%%%%%%%%%%%%%%%%%%%%%%%%%%%%%%%%%%%%% 
	\subsection{After Reheating}
	%%%%%%%%%%%%%%%%%%%%%%%%%%%%%%%%%%%%%%%%%% 
	For convenience in comparing with the results in the literature, we will first consider the regime after reheating, that is, during the radiation-dominated era, when $\arh < a < \afin$. In this case, Eq.~\eqref{eq:BE2} can be analytically solved, so that
	\begin{align} \label{eq:DNeff_RD}
		\rgw(\Tfin) &= \frac{1}{\afin^4} \int_{\arh}^{\afin} \frac{a^3}{H(a)}\, \gamma(T)\, da\, \nonumber \\
		& = \frac{3\, \mathcal{C}}{\pi}\, \sqrt{\frac{10}{\gs(\Trh)}}\, \frac{\Tfin^4\, \Trh}{M_P} \left[1 - \frac{\Tfin}{\Trh} \left[\frac{\gss(\Trh)}{\gss(\Tfin)}\right]^{4/3}\left[\frac{\gs(\Trh)}{\gs(\Tfin)}\right]^{1/2}\right] \left[\frac{\gss(\Tfin)}{\gss(\Trh)}\right]^{4/3} \nonumber\\
		& \simeq \frac{3\, \mathcal{C}}{\pi}\, \sqrt{\frac{10}{\gs(\Tfin)}}\, \frac{\Tfin^4\, \Trh}{M_P} \left[1 - \frac{\Tfin}{\Trh}\right],
	\end{align}
	where we have used Eq.~\eqref{eq:Hubble} and Eq.~\eqref{eq:Tem} in the second step. In the last step, we have assumed $\gs(\Trh) = \gs(\Tfin)$ and $\gss(\Trh) = \gss(\Tfin)$, which remains a good assumption for $\Trh \gg \Tfin \sim \mathcal{O}(100)~\text{GeV}$. With Eq.~\eqref{eq:DNeff_RD}, we find that the contribution of the GWs generated during the radiation phase using Eq.~\eqref{eq:DNeff} is
	\begin{equation} \label{eq:DNeff:RD}
		\DNeff = \DNeff^\text{RD} \simeq \frac{360\, \mathcal{C}}{7 \pi^3} \left(\frac{11}{4}\, \frac{\gss(\TDNeff)}{\gss(\Tfin)}\right)^\frac43 \sqrt{\frac{10}{\gs(\Tfin)}}\, \frac{\Trh}{M_P} \left[1 - \frac{\Tfin}{\Trh}\right].
	\end{equation}
	Since $\Tfin \ll \Trh$ is expected, $\DNeff^\text{RD} \propto \Trh/M_P$ suffers a large suppression of $1/M_P$.
	
	%%%%%%%%%%%%%%%%%%%%%%%%%%%%%%%%%%%%%%%%%%%%%%%%%%%
	\begin{figure}[t!]
		\def\sepf{0.496}
		\centering
		\includegraphics[width=\sepf\columnwidth]{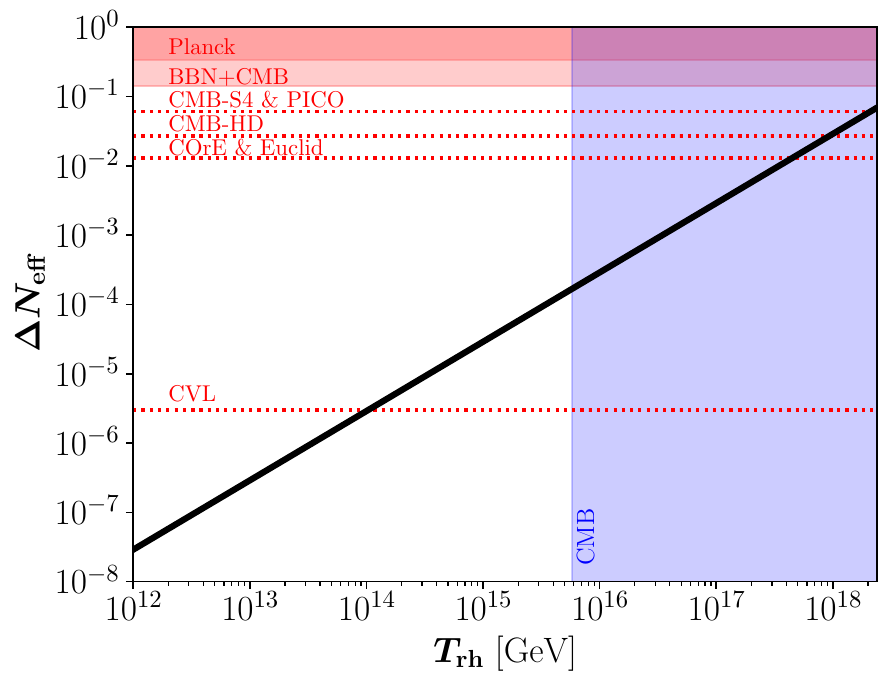}
		\caption{Contribution to $\DNeff$ as a function of $\Trh$ from the radiation-dominated era. The blue band is in tension with CMB measurements of the inflationary scale, while the red bands and red lines correspond to actual limits and future projections from Table.~\ref{tab:DNeff}.}
		\label{fig:DNeff0} 
	\end{figure} 
	%%%%%%%%%%%%%%%%%%%%%%%%%%%%%%%%%%%%%%%%%% 
	The contribution to $\DNeff$ from the radiation-dominated era is presented in Fig.~\ref{fig:DNeff0} as a function of $\Trh$, with a thick black line. Future experiments could probe large reheating temperatures close to the Planck scale, as shown by the horizontal red dotted lines (cf. Table~\ref{tab:DNeff}). However, such large values are already in tension with the maximal reheating temperature $\Trh \lesssim 5.5 \times 10^{15}$~GeV (blue vertical band), from the upper limit of BICEP/Keck on the inflationary scale~\cite{BICEP:2021xfz}.\footnote{The BICEP/Keck 2018 observations provide the most stringent limit on the inflationary tensor-to-scalar ratio, $r < 0.035$~\cite{BICEP:2021xfz}, which can be translated into a constraint on the inflationary scale: $H_I < 2 \times 10^{-5}\, M_P$. Using the fact that $\Hrh < H_I$, we derive the limit $\Trh < 5.5 \times 10^{15}$~GeV.} The proposed CVL scenario, if implemented, could, in principle, probe a significant portion of the parameter space corresponding to $\Trh \gtrsim 10^{14}$~GeV. These conclusions can be modified by including the effects of reheating on GW production, as will be discussed in the following section.
	
	%%%%%%%%%%%%%%%%%%%%%%%%%%%%%%%%%%%%%%%%%% 
	\subsection{During Reheating}
	%%%%%%%%%%%%%%%%%%%%%%%%%%%%%%%%%%%%%%%%%%
	We now compute the contribution from reheating, analytically solving Eq.~\eqref{eq:BE2} in the regime $\aini \leq a \leq \arh$. For the case where
	\begin{equation} \label{eq:boundary}
		11 - 14 \alpha + 3 \omega    
	\end{equation}
	is {\it not} zero, we obtain
	\begin{align}
		\rgw(\Trh) &  = \frac{1}{\arh^4} \int_{\aini}^{\arh} \frac{a^3}{H(a)}\, \gamma(T)\, da\, \nonumber \\
		& = \frac{6\, \mathcal{C}}{(11 - 14 \alpha + 3 \omega)\, \pi}\, \sqrt{\frac{10}{\gs(\Trh)}}\, \frac{\Trh^5}{M_P} \left[1 - \left(\frac{\Trh}{\Tini}\right)^\frac{11 - 14 \alpha + 3 \omega}{2 \alpha}\right],
	\end{align}
	and therefore 
	\begin{align} \label{eq:DNeff1}
		\DNeff &=  \DNeff^\text{RH} =\frac87 \left(\frac{11}{4}\, \frac{\gss(\TDNeff)}{\gss(\Trh)}\right)^\frac43 \frac{\gs(\Trh)}{2}\, \frac{\rgw(\Trh)}{\rR(\Trh)} \nonumber\\
		&= \frac{720\, \mathcal{C}}{7 (11 - 14 \alpha + 3 \omega)\, \pi^3} \left(\frac{11}{4}\, \frac{\gss(\TDNeff)}{\gss(\Trh)}\right)^\frac43 \sqrt{\frac{10}{\gs(\Trh)}}\, \frac{\Trh}{M_P} \left[1 - \left(\frac{\Trh}{\Tini}\right)^\frac{11 - 14 \alpha + 3 \omega}{2 \alpha}\right] \nonumber\\
		&\simeq \DNeff^\text{RD}\, \frac{2}{14 \alpha - 11 - 3 \omega} \left(\frac{\Tini}{\Trh}\right)^\frac{14 \alpha - 11 - 3 \omega}{2 \alpha} \left[1 - \left(\frac{\Trh}{\Tini}\right)^\frac{14 \alpha - 11 - 3 \omega}{2 \alpha}\right].
	\end{align}
	It is interesting to note that $\DNeff^\text{RH}$ features a power-law enhancement $(\Tini/\Trh)^\frac{14 \alpha - 11 - 3 \omega}{2 \alpha}$ with respect to the contribution during radiation domination for $14 \alpha - 11 - 3 \omega > 0$. This arises from the fact that during reheating, the temperature can be much higher than the reheating temperature $\Trh$.
	
	Alternatively, for the special case where $14 \alpha - 11 - 3 \omega = 0$, we find
	\begin{align} \label{eq:DNeff2}
		\DNeff = \DNeff^{\text{RH}} &= \frac{360\, \mathcal{C}}{7\, \alpha\, \pi^3} \left(\frac{11}{4}\, \frac{\gss(\TDNeff)}{\gss(\Trh)}\right)^\frac43 \sqrt{\frac{10}{\gs(\Trh)}}\, \frac{\Trh}{M_P}\, \ln\frac{\Tini}{\Trh} \nonumber\\
		&\simeq \DNeff^\text{RD}\, \frac{1}{\alpha}\, \ln \left(\frac{\Tini}{\Trh}\right),
	\end{align}
	with features a logarithmic boost $\ln(\Tini/\Trh)$ with respect to the contribution during radiation domination. Referring to Fig.~\ref{fig:alpha_omega}, the dashed blue line corresponds to $14 \alpha - 11 - 3 \omega = 0$. Above, at, and below this line, $\Delta N_\text{eff}$ features a power-law, logarithmic, or order-one enhancement, respectively.
	
	%%%%%%%%%%%%%%%%%%%%%%%%%%%%%%%%%%%%%%%%%%%%%%%%%%%
	\begin{figure}[t!]
		\def\sepf{0.496}
		\centering
		\includegraphics[width=\sepf\columnwidth]{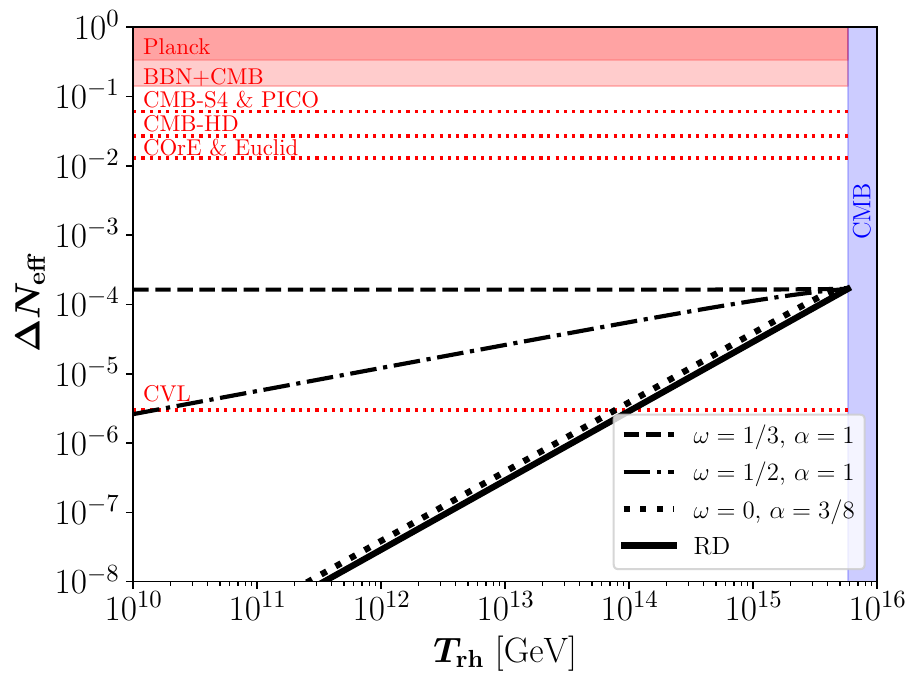}
		\caption{Total $\DNeff$ as a function of $\Trh$, summing the contributions during and after reheating, and maximizing $\Tini$ as in Eq.~\eqref{eq:Tmax}. The dashed line corresponds to $\omega = 1/3$ and $\alpha = 1$, the dot-dashed line corresponds to $\omega = 1/2$ and $\alpha = 1$, while the the dotted line to $\omega = 0$ and $\alpha = 3/8$. As a reference, the solid line (cf. Fig.~\ref{fig:DNeff0}) depicts the contribution fo the radiation-dominated era exclusively. The blue band is in tension with CMB measurements of the inflationary scale, while the red bands and red lines correspond to actual limits and future projections from Table~\ref{tab:DNeff}.}
		\label{fig:DNeff1}
	\end{figure} 
	%%%%%%%%%%%%%%%%%%%%%%%%%%%%%%%%%%%%%%%%%% 
	To show the enhancement, in Fig.~\ref{fig:DNeff1}, we illustrate $\DNeff$ as a function of the reheating temperature $\Trh$. The solid black line indicates the contribution of GWs generated solely during the radiation era (that is, Eq.~\eqref{eq:DNeff:RD}), and corresponds to the curve in Fig.~\ref{fig:DNeff0}. The additional dashed, dot-dashed and dotted curves account for the effect of the reheating phase in various combinations of $\omega$ and $\alpha$, and correspond to the {\it total} sum of contributions during and after reheating, that is, Eqs.~\eqref{eq:DNeff:RD} and~\eqref{eq:DNeff1} or~\eqref{eq:DNeff2}. For these curves, the initial temperature $\Tini$ has been {\it maximized} respecting the limits on the inflationary scale $H_I$, that is~\cite{Bernal:2024yhu}
	\begin{equation} \label{eq:Tmax}
		\Tini \leq \Trh \left[\frac{90}{\pi^2\, \gs}\, \frac{H_I^2\, M_P^2}{\Trh^4}\right]^\frac{\alpha}{3(1+\omega)}.
	\end{equation}
	In that sense, the dashed, dot-dashed, and dotted lines have to be taken as upper limits of the contribution to $\DNeff$; the solid line being the lower limit of $\DNeff$.
	
	As a first comment on Fig.~\ref{fig:DNeff1}, we note that the constraint on the inflationary scale implies that $\DNeff \lesssim \mathcal{O}(10^{-4})$.
	For the cases where $\omega = 1/3$ with $\alpha = 1$ (dashed line) and $\omega = 1/2$ with $\alpha = 1$ (dot-dashed line), $\DNeff$ has a large power-law enhancement of $(\Tini/\Trh)^1$ and $(\Tini/\Trh)^{3/4}$, respectively. However, in the case where $\omega = 0$ and $\alpha = 3/8$, there is only a small boost of $\sim 31/23$; see Eq.~\eqref{eq:DNeff1}. Furthermore, it is interesting to observe that, in the case where $\alpha = 1$ and $\omega = 1/3$, the {\it maximal} initial temperature $\Tini$ becomes independent of $\Trh$; see Eq.~\eqref{eq:Tmax}. Consequently, the expression for the maximal $\DNeff$ simplifies to $\DNeff \propto \DNeff^\text{RD} /\Trh \propto \Trh^0$. Therefore, in this scenario, the maximal $\DNeff$ is expected to be independent of $\Trh$, as shown by the dashed line in Fig.~\ref{fig:DNeff1}.
	
	%%%%%%%%%%%%%%%%%%%%%%%%%%%%%%%%%%%%%%%%%%%%%%%%%%%%%%%%%%
	\section{Gravitational Wave Spectrum} \label{sec:TGW}
	%%%%%%%%%%%%%%%%%%%%%%%%%%%%%%%%%%%%%%%%%%%%%%%%%%%%%%%%%%
	In the previous section, we computed the integrated GW spectrum and its contribution to $\DNeff$. Now we will analyze the differential spectrum. The primordial GW spectrum at present, $\Omega_{\text{GW}}(f)$, per logarithmic frequency $f$, is defined as
	\begin{equation} \label{eq:oGW}
		\Omega_{\text{GW}}(f) \equiv \frac{1}{\rho_c}\, \frac{d\rgw}{d\ln f} = \Omega_\gamma^{0}\, \frac{d(\rgw /\rR)}{d\ln k} \,,
	\end{equation}
	where the frequency relates with the momentum as $k = 2 \pi\, f$, $\rho_c$ corresponds to the critical energy density of the Universe, and $\Omega_\gamma^0\, h^2 \simeq 2.47 \times 10^{-5}$ is the photon abundance at present~\cite{Planck:2018vyg}.  
	Our next task is to obtain the differential spectrum $\frac{d(\rgw /\rR)}{d\ln k}$. To incorporate the reheating phase, it is convenient to rewrite Eq.~\eqref{eq:BE0} as
	\begin{equation} \label{eq:BEgw}
		\frac{d}{da} \left[a^4\, \frac{d\rgw}{d\ln k}\right] = \frac{a^3}{H}\, k\, \frac{d\gamma}{dk}(T, \hat k)\,,
	\end{equation}
	with which we find that the differential GW energy density at $\afin$ is 
	\begin{align}\label{eq:TGW}
		\frac{d\rgw}{d\ln\kfin}(\afin) &= \frac{1}{\afin^4} \Bigg[\int_{\aini}^{\arh} da\, \frac{a^3\, k(a)}{H(a)}\, \frac{d\gamma}{dk}\left(T(a), \hat k(a)\right) + \int_{\arh}^{\afin} da\, \frac{a^3\, k(a)}{H(a)}\, \frac{d\gamma}{dk}\left(T(a), \hat k(a)\right)\Bigg] \nonumber\\
		&\simeq \frac{\krh}{\afin^4\, \arh\, \Hrh} \Bigg[\int_{\aini}^{\arh} da\, a^4 \left(\frac{\arh}{a}\right)^\frac{1-3\omega}{2} \frac{d\gamma}{dk}\left(\Trh \left(\frac{\arh}{a}\right)^\alpha, \hat k_\text{rh} \left(\frac{a}{\arh}\right)^{\alpha-1}\right) \nonumber\\
		&\qquad\qquad\qquad\quad + \int_{\arh}^{\afin} da\, a^4\, \frac{d\gamma}{dk}\left(\Trh \frac{\arh}{a}, \hat k_\text{fin}\right)\Bigg] \nonumber\\
		&\simeq \frac{2\, \hat k_\text{fin}^3\, \Tfin^3\, \Trh^4}{\pi^2\, M_P^2\, \Hrh\, \afin} \Bigg[ \int_{\aini}^{\arh} da \left(\frac{\arh}{a}\right)^\frac{8\alpha - 3\omega - 3}{2} \hat\eta\left(\hat k_\text{rh} \left(\frac{a}{\arh}\right)^{\alpha-1}\right) \nonumber\\
		&\qquad\qquad\qquad + \hat\eta(\hat k_\text{fin}) \int_{\arh}^{\afin} da  \left(\frac{\arh}{a}\right)^2\Bigg],
	\end{align}
	by using Eqs.~\eqref{eq:Hubble} and~\eqref{eq:Tem}, and noticing that the graviton momenta always scales as $k(a) \propto 1/a$ regardless of the behavior of the background. The first term inside the square brackets represents the contribution from the reheating phase, while the second is the contribution from the radiation era. In the second part, the running of the couplings (and thus the temperature dependence of the source term) has been neglected. In the next sections, the two terms will be independently analyzed.
	
	%%%%%%%%%%%%%%%%%%%%%%%%%%%%%%%%%%%%%
	\subsection{After Reheating}
\begin{figure}[t!]
		\def\sepf{0.496}
		\centering
		\includegraphics[width=\sepf\columnwidth]{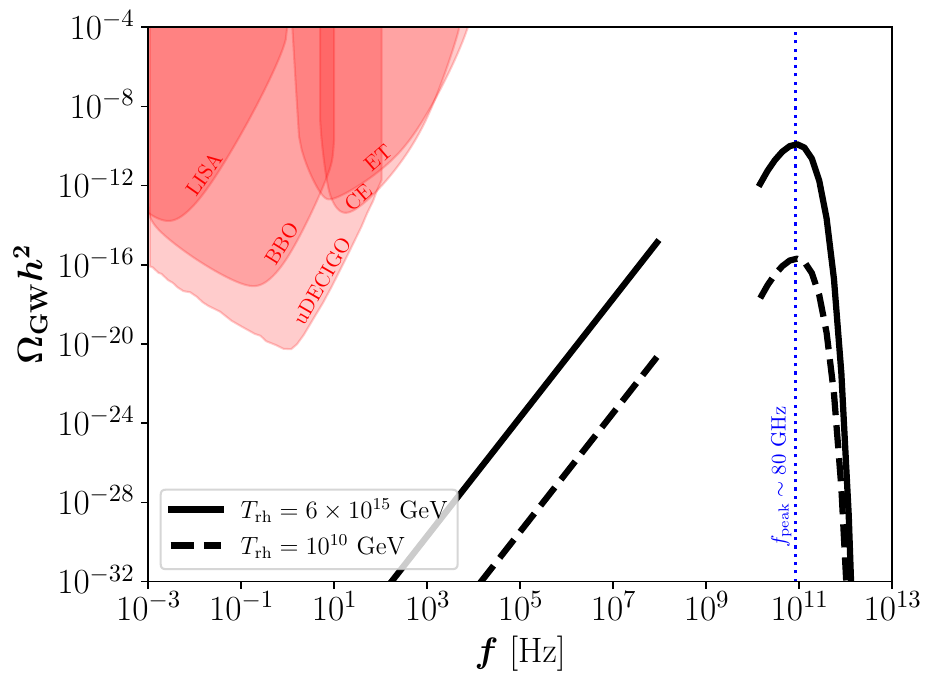}
		\caption{GW spectrum produced in the radiation-dominated era, for different $\Trh$ (black lines). The vertical blue dotted line corresponds to the peak frequency $\fpeak \sim 80$~GHz. The sensitivity curves of several proposed GW detectors are shown in red.}
		\label{fig:GW_RD}
	\end{figure}

	%%%%%%%%%%%%%%%%%%%%%%%%%%%%%%%%%%%%%
	We first focus on the contribution after reheating, in the radiation-dominated phase, namely the second term in Eq.~\eqref{eq:TGW}. We find that
	\begin{align}\label{eq:diff_RD}
		\frac{d\rgw}{d\ln\hat k_\text{fin}}(\afin) &\simeq \frac{2\, \hat k_\text{fin}^3\, \Tfin^3\, \Trh^4\, \hat\eta(\hat k_\text{fin})}{\pi^2\, M_P^2\, \Hrh\, \afin} \int_{\arh}^{\afin} da \left(\frac{\arh}{a}\right)^2 \simeq \frac{2\, \hat k_\text{fin}^3\, \Tfin^3\, \Trh^4\, \hat\eta(\hat k_\text{fin})}{\pi^2\, M_P^2\, \Hrh}\, \frac{\arh}{\afin} \left[1-\frac{\arh}{\afin}\right] \nonumber\\
		&\simeq \frac{2\, \hat k_\text{fin}^3\, \Tfin^4\, \Trh^3\, \hat\eta(\hat k_\text{fin})}{\pi^2\, M_P^2\, \Hrh} \left[1 - \frac{\Tfin}{\Trh}\right] \simeq \frac{6}{\pi^3} \sqrt{\frac{10}{\gs(\Trh)}}\, \frac{\Tfin^4\, \Trh}{M_P}\, \hat k_\text{fin}^3\, \hat\eta(\hat k_\text{fin})\,,
	\end{align}
	with which we find that at present
	\begin{equation}
		\frac{d\rgw}{d\ln\hat k_\text{fin}}(a_0) = \frac{d\rgw}{d\ln\hat k_\text{fin}}(\afin) \left(\frac{\afin}{a_0}\right)^4 \simeq \frac{6}{\pi^3} \left[\frac{\gss(T_0)}{\gss(\Tfin)}\right]^\frac43 \sqrt{\frac{10}{\gs(\Trh)}}\, \frac{T_0^4\, \Trh}{M_P}\, \hat k_\text{fin}^3\, \hat\eta(\hat k_\text{fin})\,.
	\end{equation}
	Using Eq.~\eqref{eq:oGW}, we obtain the GW spectrum from the radiation-dominated era
	\begin{equation}
		\Omega^\text{RD}_\text{GW}(f) = \Omega^0_\gamma\, \frac{d(\rgw/\rR)}{d\ln \hat k_\text{fin}}(a_0) \simeq \Omega^0_\gamma\, \frac{18}{\pi^5} \left[\frac{\gss(T_0)}{\gss(\Tfin)}\right]^\frac43 \sqrt{\frac{10}{\gs(\Trh)}}\, \frac{10}{\gs(T_0)}\, \frac{\Trh}{M_P}\, \hat k_\text{fin}^3\, \hat\eta(\hat k_\text{fin})\,,
	\end{equation}
	which is linearly proportional to $\Trh$~\cite{Ghiglieri:2020mhm, Ringwald:2020ist}. 
	The frequency $f$ relates to $\hat k_\text{fin}$ as
	\begin{equation}
		\hat{k}_\text{fin} = \frac{k_\text{fin}}{T_\text{fin}} = \frac{k_0}{T_0} \left[\frac{\gss(T_\text{fin})}{\gss(T_0)}\right]^{1/3} = \frac{2 \pi\, f }{T_0} \left[\frac{\gss(T_\text{fin})}{\gss(T_0)}\right]^{1/3},
	\end{equation}
	with which we find the peak frequency is located around
	\begin{equation}
		\fpeak \simeq  \frac{3.92}{2\pi}\left[\frac{\gss (T_0)}{\gss(\Tfin)}\right]^{1/3}  T_0 \simeq 74~\text{GHz}\,,
	\end{equation}
	independently of the reheating temperature. We note that the factor 3.92 corresponds to the position of the maximum of the function $\hat k\, \frac{d\gamma}{d\hat k} \sim \hat k^3\, \hat\eta(\hat k) \sim \hat k^4/(e^{\hat k}-1)$ in Eq.~\eqref{eq:BEgw}, neglecting the logarithmic part of the function $\hat\eta$ in Eq.~\eqref{eq:HTL}. Instead, if the logarithmic term is taken into account, we find that the peak is $\fpeak \simeq 80$~GHz, as shown in Fig.~\ref{fig:GW_RD}.
	%%%%%%%%%%%%%%%%%%%%%%%%%%%%%%%%%%%%%%%%%%%%%%%%%%%
	%%%%%%%%%%%%%%%%%%%%%%%%%%%%%%%%%%%%%%%%%%
	
	In Fig.~\ref{fig:GW_RD}, we present the GW spectrum from the radiation-dominated era. For comparison, we also include the sensitivity bands of several proposed GW detectors, such as LISA~\cite{LISA:2017pwj}, the Einstein Telescope (ET)~\cite{Punturo:2010zz, Hild:2010id, Sathyaprakash:2012jk, Maggiore:2019uih}, the Big Bang Observer (BBO)~\cite{Crowder:2005nr, Corbin:2005ny, Harry:2006fi}, and the ultimate DECIGO (uDECIGO)~\cite{Seto:2001qf, Kudoh:2005as}, and the Cosmic Explorer (CE)~\cite{Reitze:2019iox}. For development on ultrahigh-frequency GWs, see Ref.~\cite{Aggarwal:2020olq} for a recent review. We consider two benchmark scenarios with $\Trh = 6 \times 10^{15}$~GeV (solid black line) and $\Trh = 10^{10}$~GeV (dashed black line). For the highest allowed reheating temperature, $\Trh = 6 \times 10^{15}$~GeV, the peak amplitude reaches $\Omega^\text{RD}_\text{GW}(f_{\text{peak}}) \simeq 10^{-10}$, which remains well beyond the sensitivity of current experimental constraints.\footnote{The energy stored in GWs behaves like dark radiation, contributing to the effective number of neutrino species. For a given $\DNeff$, $\Omega_\text{GW} h^2 \lesssim 5.6 \times 10^{-6}\, \DNeff$~\cite{Caprini:2018mtu}.} As we will demonstrate in the next section, the amplitude can be significantly enhanced when the reheating phase is included in the analysis.
	
	%%%%%%%%%%%%%%%%%%%%%%%%%%%%%%%%%%%%%
	\subsection{During Reheating}
	%%%%%%%%%%%%%%%%%%%%%%%%%%%%%%%%%%%%%
	We now focus on the contribution to the GW spectrum during reheating, namely, the first term of Eq.~\eqref{eq:TGW}, which is 
	\begin{equation}\label{eq:diff_RH}
		\frac{d\rgw}{d\ln\kfin}(\afin) \simeq \frac{2\, \hat k_\text{fin}^3\, \Tfin^3\, \Trh^4}{\pi^2\, M_P^2\, \Hrh\, \afin} \int_{\aini}^{\arh} da \left(\frac{\arh}{a}\right)^\frac{8\alpha - 3\omega - 3}{2} \hat\eta\left(\hat k_\text{rh} \left(\frac{a}{\arh}\right)^{\alpha-1}\right).
	\end{equation}
	Introducing the dimensionless variable $x \equiv a/\arh$, and comparing with Eq.~\eqref{eq:diff_RD}, one has that
	\begin{equation} \label{eq:OGW_RH}
		\Omega^{\text{RH}}_\text{GW}(f) \simeq \Omega^{\text{RD}}_\text{GW}(f) \times \left[ \frac{1}{\hat\eta^{\text{RD}}(\hat k_\text{rh})}\int_{\aini/\arh}^{1} dx\, \left(\frac{1}{x}\right)^\frac{8\alpha-3\omega -3 }{2}\, \hat\eta\left(\hat k_\text{rh} \,x^{\alpha-1}\right) \right],
	\end{equation}
	where the term within the square bracket corresponds to a potential enhancement, which depends on $\Tini/\Trh$ (via the integration limit $\aini/\arh$), $\alpha$, and $\omega$. For instantaneous reheating, the term within the square brackets approaches 1, and there is no enhancement.  We also note that the bulk of the contribution occurs when $x \ll 1$, as expected, since graviton production is a UV-dominated freeze-in process.  
	
	During reheating, the temperature scales as $T(a) \propto a^{-\alpha}$, which can differ from the scaling of the graviton momentum, $k \propto a^{-1}$. As a result, the analytical integration of the source term becomes, in general, very challenging for $\alpha \neq 1$.\footnote{The analytical solution can be written in terms of a difference of incomplete gamma functions, not particularly enlightening for ordinary humans.} For the special case with $\alpha = 1$, we find
	\begin{align} \label{eq:OGW_RH_special}
		\Omega^{\text{RH}}_\text{GW}(f) & \simeq \Omega^{\text{RD}}_\text{GW}(f) \times \int_{\aini/\arh}^{1} dx\, x^\frac{3\omega - 5}{2} \nonumber \\
		& = \Omega^{\text{RD}}_\text{GW}(f)\times
		\begin{dcases}
			\frac{2}{3(1-\omega)} \left[\left(\frac{\Tini}{\Trh} \right)^{\frac{3(1-\omega)}{2}} -1\right] & \text{for $\omega \neq 1$}\,, \\
			\log \left(\frac{\Tini}{\Trh} \right) & \text{for $\omega = 1$}\,,
		\end{dcases}
	\end{align}
	which features a power-law enhancement $\sim (\Tini/\Trh)^{3(1-\omega)/2}$ compared to the pure radiation case for $\omega \neq 1$.  We also note that in this case where $\alpha = 1$, the location of the peak is the same as in the radiation-dominated scenario. 
	
	%%%%%%%%%%%%%%%%%%%%%%%%%%%%%%%%%%%%%%%%%%%%%%%%%%%
	\begin{figure}[t!]
		\def\sepf{0.45}
		\centering    
		\includegraphics[width=\sepf\columnwidth]{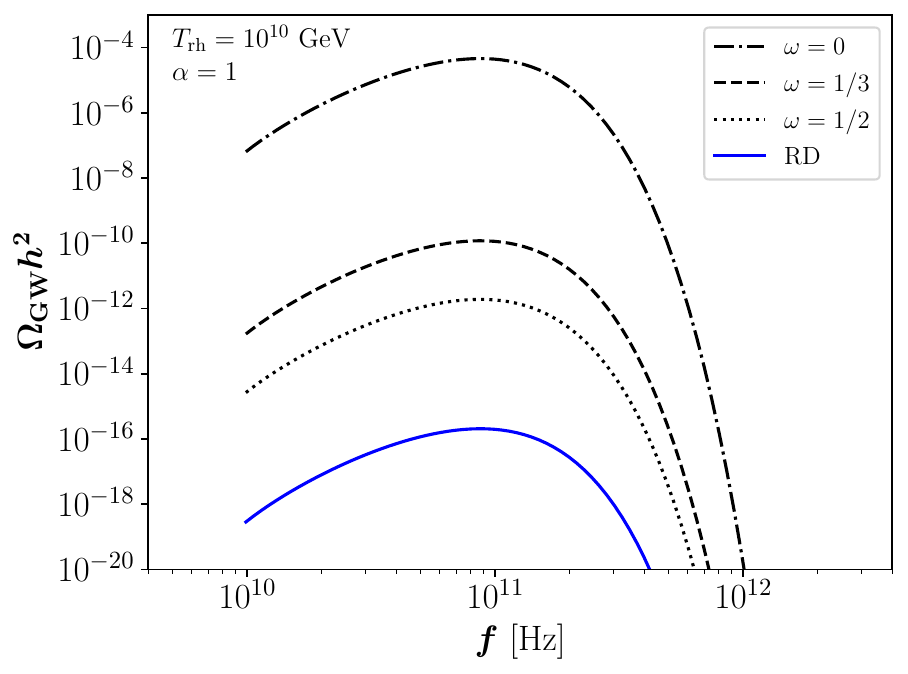}
		\includegraphics[width=\sepf\columnwidth]{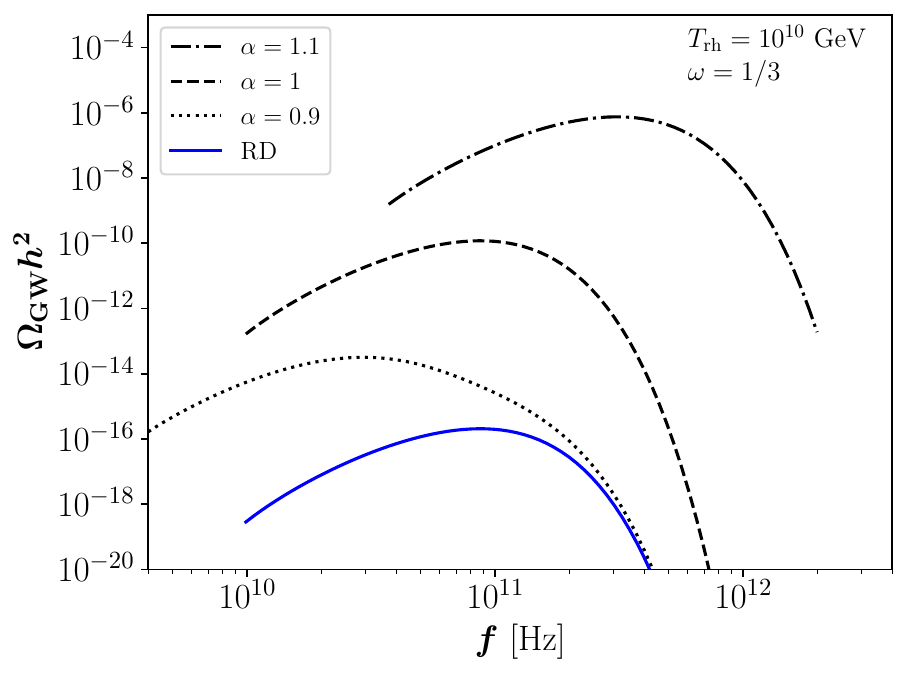}
		\includegraphics[width=\sepf\columnwidth]{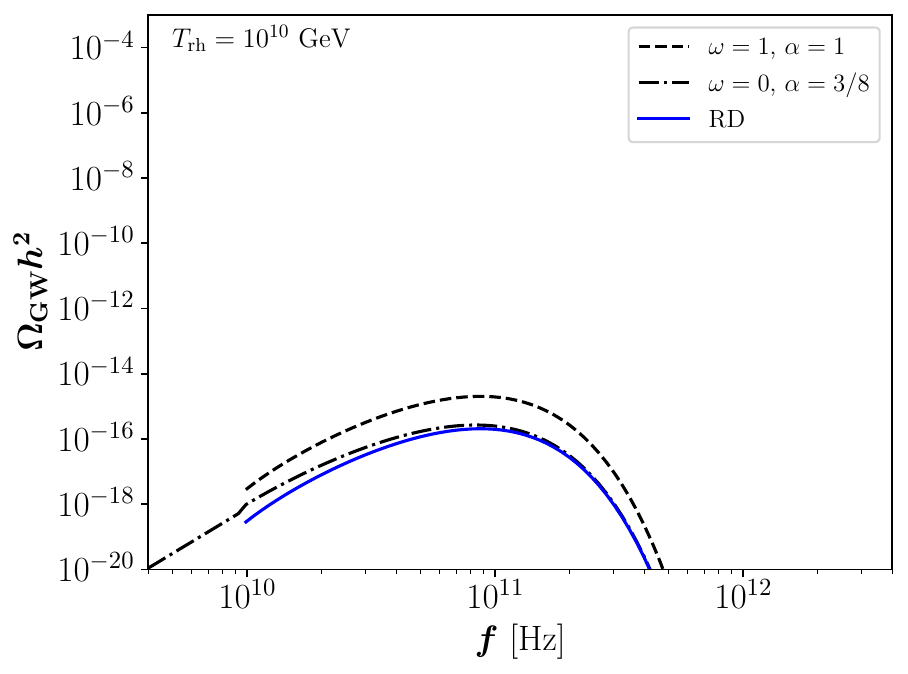}
		\caption{GW spectra for $\Trh = 10^{10}$~GeV. The blue solid lines (RD) correspond to the contribution after reheating, while the black lines to the full spectrum (during and after reheating), for different reheating scenarios. Top left: $\alpha=1$ with $\omega = 0$, 1/3 or 1/2. Top right: $\omega=1/2$ with $\alpha = 1.1$, 1 or 0.9. Bottom: $\omega = \alpha = 1$ and $\omega = 0$ with $\alpha=3/8$.}
		\label{fig:GW_RH}
	\end{figure} 
	%%%%%%%%%%%%%%%%%%%%%%%%%%%%%%%%%%%%%%%%%%
	In Fig.~\ref{fig:GW_RH}, fully numerical solutions of the GW spectrum are shown, for $\Trh = 10^{10}$~GeV and different combinations of $\omega$ and $\alpha$. The blue lines correspond to the contribution from the radiation-dominated era, while the black lines correspond to the total spectrum with the contribution from reheating. In that case, the maximal inflationary scale has been used to compute $\Tini$.
	\begin{itemize}
		\item In the upper left panel of Fig.~\ref{fig:GW_RH}, we investigate the impact of $\omega$ on the GW spectrum. To this end, we fix $\alpha = 1$, and consider $\omega = 0$ (dot-dashed black line), $\omega = 1/3$ (dashed black line), and $\omega = 1/2$ (dotted black line). In this scenario, the peak frequency remains the same as in the pure radiation case, around $\fpeak \simeq 80$~GHz. Compared to the pure radiation scenario (blue line), we find a significant enhancement in the GW amplitude, as also suggested by the first line of Eq.~\eqref{eq:OGW_RH_special}. Moreover, as $\omega$ increases, the enhancement decreases because the maximum temperature $\Tini$ decreases, leading to a suppression of graviton production during reheating.
		\item In the upper right panel of Fig.~\ref{fig:GW_RH}, we explore the effect of $\alpha$ on the GW spectrum. We fix $\omega = 1/3$ and vary $\alpha$, taking $\alpha = 1.1$ (dot-dashed black line), $\alpha = 1$ (dashed black line) and $\alpha = 0.9$ (dotted black line). Unlike in the previous case, the peak frequency shifts to high frequencies as $\alpha$ increases. This is because a larger $\alpha$ corresponds to a higher $\Tini$ (cf. Eq.~\eqref{eq:Tmax}), implying that the bulk of the production occurs much earlier, hence with higher frequencies. Similarly for the amplitude, we find that a higher value of $\alpha$ leads to a more pronounced enhancement in the GW spectrum. 
		\item Finally, we explore two special cases in the lower panel of Fig.~\ref{fig:GW_RH}  with $\omega = \alpha = 1$ and $\omega = 0$ with $\alpha = 3/8$. The former case corresponds to the second line of Eq.~\eqref{eq:OGW_RH_special}, where a mild logarithmic enhancement is expected. In fact, numerically, we find an enhancement factor $\sim 10$, consistent with the analytical estimation. The latter case corresponds to the scenario where the reheating phase is governed by a non-relativistic matter field. In such a case, we find only an order-one enhancement. These features are similar to the boost on $\DNeff$; see Eqs.~\eqref{eq:DNeff1} and~\eqref{eq:DNeff2}.
	\end{itemize}
	
	%%%%%%%%%%%%%%%%%%%%%%%%%%%%%%%%%%%%%%%%%%%%%%%%%%%
	\begin{figure}[t!]
		\def\sepf{0.496}
		\centering
		\includegraphics[width=\sepf\columnwidth]{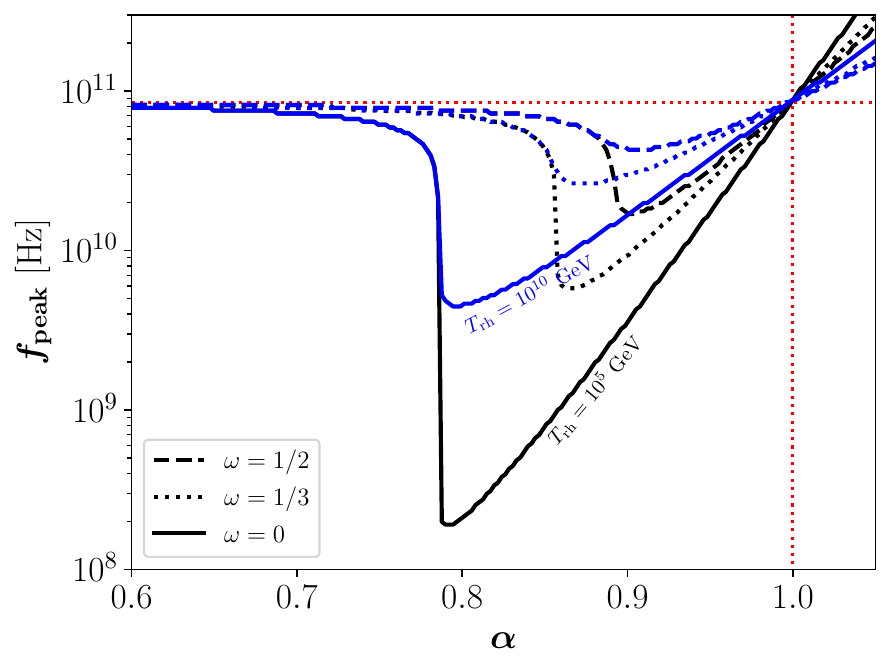}
		\includegraphics[width=\sepf\columnwidth]{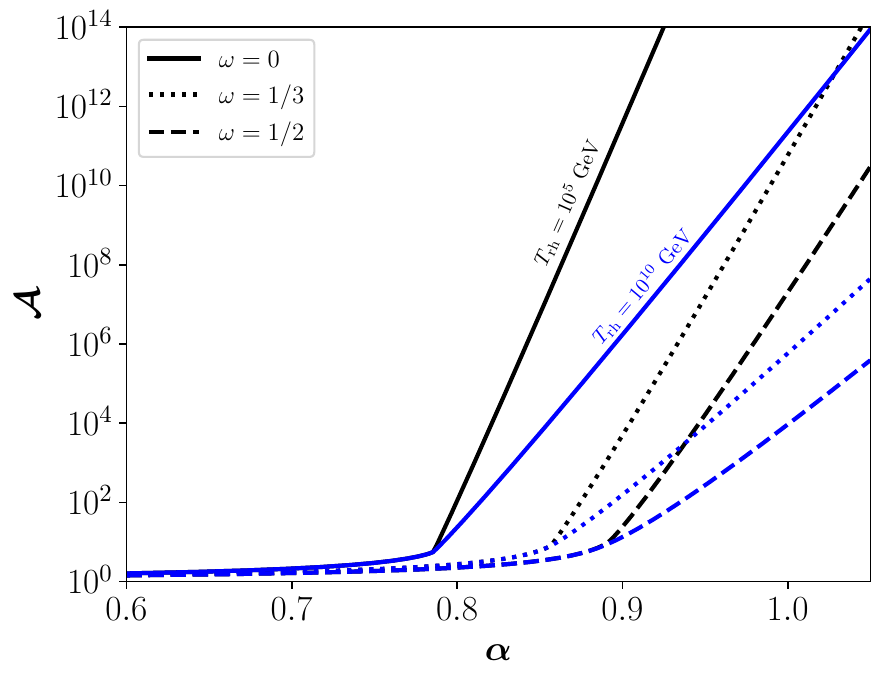}
		\includegraphics[width=\sepf\columnwidth]{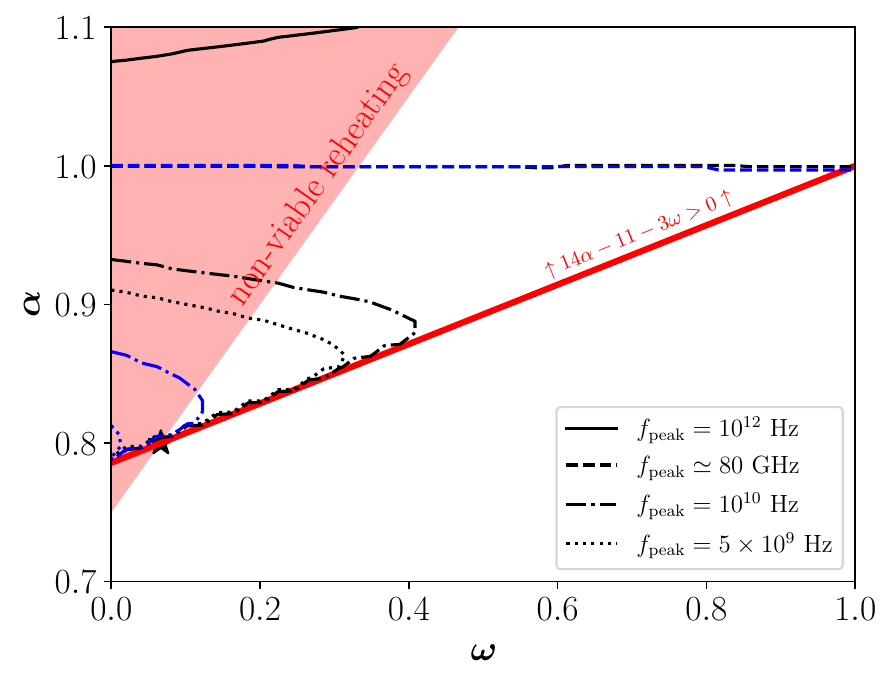}
		\includegraphics[width=\sepf\columnwidth]{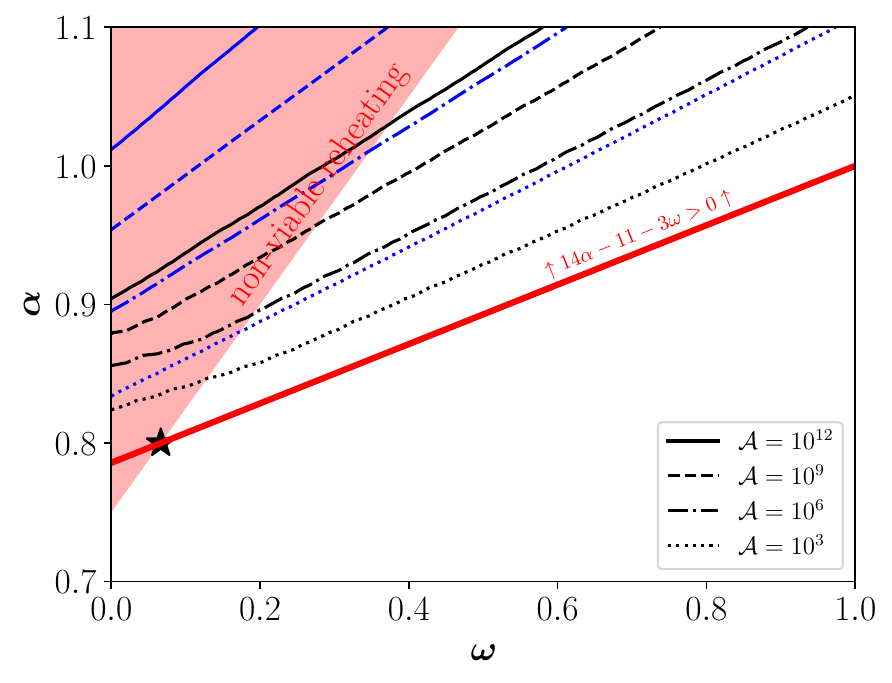}
		\caption{Left panels: Frequency $\fpeak$ at which the GW spectrum reaches its maximum, for $\Trh = 10^5$~GeV (black) and $\Trh = 10^{10}$~GeV (blue). Right panels: Amplification of the spectrum, defined as the maximal amplitude of the full spectrum (at $f=\fpeak$) over the maximal amplitude of the spectrum with only the contribution from the radiation-dominated era. Lower panels: The red area in the upper left corner does not give rise to viable reheating. Above, over and below the solid red line, the GW spectrum has a power-law, logarithmic or order-one enhancement during reheating, respectively; cf. Eq~\eqref{eq:DNeff1}. }
		\label{fig:peak}
	\end{figure}
\begin{figure}[t!]
		\def\sepf{0.496}
		\centering
		\includegraphics[width=\sepf\columnwidth]{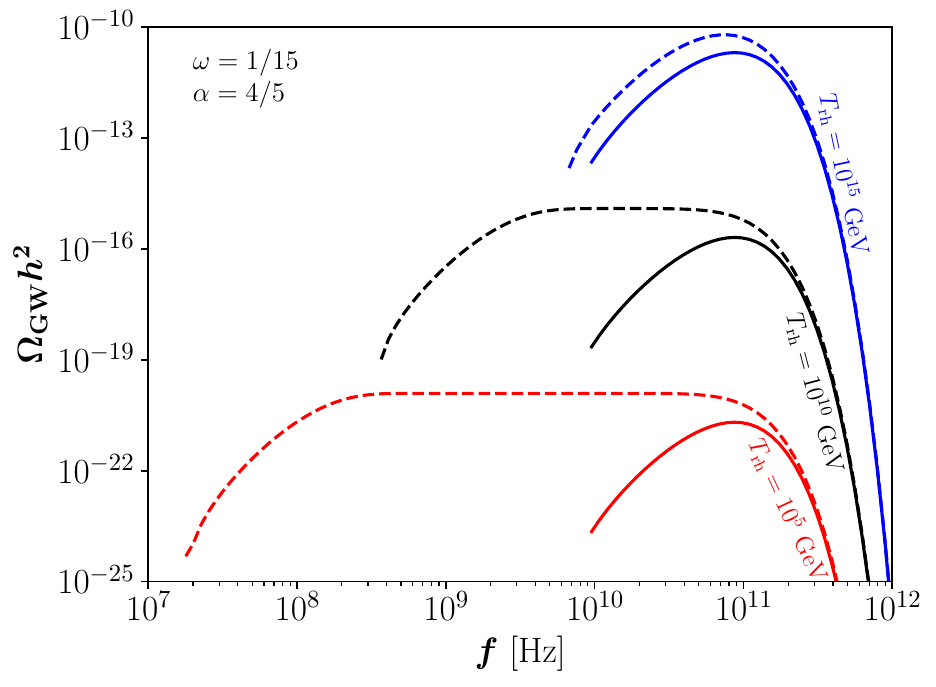}
		\caption{GW spectra for $\Trh = 10^{15}$~GeV (blue), $10^{10}$~GeV (black) and $10^5$~GeV (red). The solid lines correspond to the contribution after reheating, while the dashed lines to the full spectrum (during and after reheating) for $\omega = 1/15$ and $\alpha = 4/5$.}
		\label{fig:plateau}
	\end{figure}

 
	%%%%%%%%%%%%%%%%%%%%%%%%%%%%%%%%%%%%%%%%%%
	As noted above, the location of the peak and the amplitude vary with $\alpha$ and $\omega$. In Fig.~\ref{fig:peak}, we illustrate these features in more detail. The blue lines represent $\Trh = 10^5$~GeV, while the black lines correspond to $\Trh = 10^{10}$~GeV. We have again fixed the inflationary scale parameter in Eq.~\eqref{eq:Tmax} to its maximum value. 
	\begin{itemize}
		\item In the upper-left panel of Fig.~\ref{fig:peak}, we examine the relationship between the peak frequency $\fpeak$ and $\alpha$.  Solid, dotted and dashed lines indicate $\omega = 0$, $\omega = 1/3$, and $\omega = 1/2$, respectively.  The horizontal dotted red line corresponds to $\fpeak \simeq 80$~GHz, which corresponds to the peak frequency for GW in radiation domination. Taking reheating into account, we find that $\fpeak$ approaches the horizontal dotted red line for $\alpha = 1$, and for small values of $\alpha < (11 + 3\omega)/14$, where there is no significant enhancement of the spectrum; cf. Eq.~\eqref{eq:boundary}. In the range $(11+3\omega)/14 < \alpha < 1$, the peak shifts to smaller frequencies, with larger displacements occurring in the case of small values of $\Trh$, that is, long reheating periods, where the GWs experience more redshift.
		\item In the upper right panel of Fig.~\ref{fig:peak}, we explore the amplification of the spectrum, defined as the maximal amplitude of the full spectrum (at $f=\fpeak$) over the maximal amplitude of the spectrum with only the contribution from the radiation-dominated era, namely
		\begin{equation} \label{eq:A}
			\mathcal{A} = \frac{\max\left[\Omega^\text{RD}_\text{GW} + \Omega^\text{RH}_\text{GW}\right]}{\max\left[ \Omega^{\text{RD}}_\text{GW}\right]}\,.
		\end{equation}
		For $\alpha < (11+3\omega)/14$ there is an order-one boost (corresponding to the flat part of the curves), whereas for $\alpha > (11+3\omega)/14$ the maximum of the spectrum has a power-law amplification. For fixed values of $\omega$ and $\alpha$, a smaller $\Trh$ implies larger $\Tini$ (cf. Eq.~\eqref{eq:Tmax}), resulting in greater amplification. This explains the slopes for the blue lines with respect to the black lines. 
		\item The behavior of $\fpeak$ is further studied in the lower left panel of Fig.~\ref{fig:peak}, now in the plane $[\omega,\, \alpha]$. The solid, dashed, dot-dashed, and dotted lines correspond to $\fpeak = 10^3$~GHz, $80$~GHz, $10^1$~GHz, and $5$~GHz, respectively.
		$(i)$ Below the red line  with $14 \alpha -11 -3\omega <0$, the enhancement of the GW spectrum is insignificant and therefore the peak stays near 80~GHz.
		$(ii)$ Along the red line corresponding to $14 \alpha -11 -3\omega =0$, there is no enhancement for the spectrum, i.e. the spectrum features a plateau, that will be further explored in Fig.~\ref{fig:plateau}. This feature can also be seen by vertical lines in the upper left panel. We note that  such plateau feature is maximized for the lowest values of $\alpha$ compatible with viable reheating, that is, $\omega = 1/15$ and $\alpha = 4/5$ (black star {\boldmath $\star$} in the lower panels of Fig.~\ref{fig:peak}). 
		$(iii)$  Above the red line with $14 \alpha -11 -3\omega >0$, there are three regime for $\alpha$. For $\alpha <1$, it is necessary that $\alpha$ decreases with $\omega$ to yield a fixed $\fpeak$ as shown by the dash dotted and dotted lines. In the regime with $\alpha >1$,  $\alpha$ increases with $\omega$ as shown in the solid line. Finally, for $\alpha=1$, changing reheating temperature would only modify the amplitude of the peak, not the location. This is why the black dashed and blue dashed lines overlap with each other. These features can also be seen in the upper-left panel of Fig.~\ref{fig:peak}.
		\item Finally, the lower-right panel of Fig.~\ref{fig:peak} explores the amplification factor $\mathcal{A}$, defined in Eq.~\eqref{eq:A}, as a function of $\omega$ and $\alpha$. The solid, dashed, dot-dashed, and dotted lines correspond to $\mathcal{A} = 10^{12}$, $\mathcal{A} = 10^9$, $\mathcal{A} = 10^6$, and $\mathcal{A} = 10^3$, respectively. We remind the reader that the black lines represent $\Trh = 10^5$~GeV, while the blue lines correspond to $\Trh = 10^{10}$~GeV. Focusing first on $\Trh = 10^5$~GeV, as mentioned above, the further away the curves are from the red line, the larger $\mathcal{A}$ becomes. This explains why the solid black line with $\mathcal{A} = 10^{12}$ is much further from the red line compared to the other three black curves. For the blue curves with larger $\Trh = 10^{10}$~GeV (i.e., a larger reheating temperature), it is expected that the amplification factor $\mathcal{A}$ is smaller (cf. upper-right panel). This implies that to achieve a fixed amplification factor, such as $\mathcal{A} = 10^3$, a larger $\alpha$ is needed to compensate. This explains why the blue lines are above the black lines for the same $\Trh$ at a fixed $\mathcal{A}$.
	\end{itemize}
	
	%%%%%%%%%%%%%%%%%%%%%%%%%%%%%%%%%%%%%%%%%%%%%%%%%%%
	%%%%%%%%%%%%%%%%%%%%%%%%%%%%%%%%%%%%%%%%%%
	To demonstrate the novel plateau feature mentioned above, Fig.~\ref{fig:plateau} shows the GW spectrum for the case $\omega = 1/15$ with $\alpha = 4/5$  for $\Trh = 10^{15}$~GeV (blue), $10^{10}$~GeV (black) and $10^5$~GeV (red). The solid lines correspond to the contribution after reheating, whereas the dashed lines correspond to the full spectrum (during and after reheating). A prominent plateau  spanning from $f \sim 80$~GHz to lower frequencies can be seen. Even a small change in $\alpha$ tilts the plateau, generating a maximum near $f \sim 80$~GHz or at smaller frequencies, if $\alpha$ is reduced or increased, respectively, explaining the fast transition in the upper left panel of Fig.~\ref{fig:peak}. 
	
	Before closing this section, we emphasize that the production of GW during reheating not only boosts the amplitude of the GW spectrum but also can alter its shape, introducing new spectral features. 
	
	%%%%%%%%%%%%%%%%%%%%%%%%%%%%%%%%%%%%%%%%%%%%%%%%%%%%%%%%%%
	\section{Conclusions} \label{sec:concl}
	%%%%%%%%%%%%%%%%%%%%%%%%%%%%%%%%%%%%%%%%%%%%%%%%%%%%%%%%%%
	In this study, we revisited the production of the stochastic gravitational wave (GW) background generated through graviton emission from interactions of standard model (SM) particles in the thermal plasma. The GW spectrum produced during the radiation-dominated era peaks at a frequency of approximately $80$~GHz, with an amplitude that scales linearly with the reheating temperature.
	
	We extended existing results in the literature by incorporating the cosmic reheating phase in the thermal history of the Universe. Since graviton production is suppressed by powers of the Planck scale, it exhibits a UV freeze-in behavior, implying that it mainly occurs at the highest temperatures reached by the bath, typically during the reheating era. Being agnostic about the evolution of the background during reheating, we conveniently parametrize the evolution of the SM temperature and the Hubble expansion rate as a function of the cosmic scale factor $a$ as $T(a) \propto a^{-\alpha}$ and $H(a) \propto a^{-3(1+\omega)/2}$, respectively. This general parametrization allows us to explore various reheating scenarios.
	
	Compared to the case of pure radiation domination, there is a significant enhancement in GW production, with typically a strong dependence on the ratio of the maximum temperature to the reheating temperature. The boost in the total energy density in the form of GWs can be quantified by its contribution to the effective number of neutrinos $\DNeff$, as shown in Eq.~\eqref{eq:DNeff1} and Eq.~\eqref{eq:DNeff2}. We find that for $11-14 \alpha +3\omega>0$, $=0$ or $<0$, there is a power law, logarithmic, or order-one enhancement, respectively. The result is summarized in Fig.~\ref{fig:DNeff1}, where we demonstrate that the enhancement brings the $\DNeff$ contribution from the GW spectrum closer to the sensitivity of future experiments, even in the case of very low reheating temperatures.
	
	The enhancement also manifests itself in the differential GW spectrum. In the case where $11 - 14 \alpha + 3 \omega > 0$, we find that the GW spectrum exhibits a power-law enhancement, as shown in Eq.~\eqref{eq:OGW_RH_special} and Fig.~\ref{fig:GW_RH}. Furthermore, beyond the overall boost in amplitude, the GW spectrum produced during reheating can peak at different frequencies when $\alpha \ne 1$, as demonstrated in Fig.~\ref{fig:peak}. Moreover, we observe novel characteristic features, with a broader spectrum emerging, as illustrated in Fig.~\ref{fig:plateau}. 
	
	In summary, this work presents a thorough exploration of GW production from the thermal plasma, taking the reheating phase into account. We have highlighted new features of the GW spectrum compared to the pure radiation domination case.
	
	%%%%%%%%%%%%%%%%%%%%%%%%%%%%%%%%%%%%%%%%%%%
	\acknowledgments
	%%%%%%%%%%%%%%%%%%%%%%%%%%%%%%%%%%%%%%%%%%%
	NB thanks the hospitality of the Mainz Institute for Theoretical Physics (MITP) during the workshop ``The Dark Matter Landscape: From Feeble to Strong Interactions''.  The authors especially thank Nicolás Fernández for the earlier collaboration and discussions.  We also acknowledge useful conversations with Fernando Quevedo, Carlos Tamarit and Gonzalo Villa. NB received funding from the Spanish FEDER / MCIU-AEI under the grant FPA2017-84543-P. YX has received support from the Cluster of Excellence ``Precision Physics, Fundamental Interactions, and Structure of Matter'' (PRISMA$^+$ EXC 2118/1) funded by the Deutsche Forschungsgemeinschaft (DFG, German Research Foundation) within the German Excellence Strategy (Project No. 390831469).
	
	%%%%%%%%%%%%%%%%%%%%%%%%%
	\bibliographystyle{JHEP}
	\bibliography{biblio}
	%%%%%%%%%%%%%%%%%%%%%%%%%
\end{document}